\documentstyle[twoside,fleqn,espcrc2]{article}

\def\d{\partial}
\def\r{\rho}
\def\rd{{\dot \rho}}
\def\rt{{\tilde \rho}}
\def\v{\upsilon}
\def\rv{{\rm v}}
\def\l{\lambda}
\def\rs{\rho_{\rm sol}}
\def\rw{\rho_{\rm wave}}
\def\half{{\textstyle {1 \over 2}}}
\def\NP{{\it Nucl. Phys.\ }}
\def\PL{{\it Phys. Lett.\ }}

\def\PR{{\it Phys. Rev. \ }}
\def\PRL{{\it Phys. Rev. Lett.\ }}

\def\JP{{\it J. Phys.\ }}
\def\IJMP{{\it Int. Jour. Mod. Phys.\ }}

\newcommand{\AmS}{{\protect\the\textfont2
  A\kern-.1667em\lower.5ex\hbox{M}\kern-.125emS}}

\title{Solitons and fractional statistics}

\author{Alexios P. Polychronakos\address{Theoretical Physics Dept.,
        Uppsala University \\
        Box 803, S-75108 Uppsala, Sweden}%
        \thanks{Part of this work was done while still at CERN}
	}

\begin{document}

\begin{abstract}
Solitons in the continuum limit of the Calogero model are derived
and shown to correspond to one-particle excitations. The statistical
mechanics of exclusion statistics particles is then formulated in
terms of a priori probabilities and a path integral is thereoff
constructed.

\vskip 0.5cm
{\it Talk delivered at the Trieste 10-12 April 1995 Conference on
statistical mechanics and QFT, and at the Oslo 21-26 August 1995
Worskhop on low-dimensional systems.}
\end{abstract}

\maketitle

\section{INTRODUCTION}

This talk consists, in fact, of two short stories, both connected
to or motivated from the Calogero model. In the first, I derive
an analytic expression for a solitonic wave excitation in the
continuum limit of the Calogero model, and show that it corresponds
to one-particle excitations in the quantum description. Large-amplitude
waves are also derived and correspond to a two-band quantum state.
A conjecture for the chiral hamiltonian of this problem is made.
In the second, I formulate exclusion statistics in terms of
microscopic occupation probabilities. This leads to a path-integral
construction of the partition function for such systems and can
be used as a starting point for further generalizations. A bonus
of this formalism is that the coefficients of the low-temperature
expansion of the specific heat can be easily evaluated. Much of the
material in this talk is taken from references \cite{AP}.

\section{SOLITONS}

The Calogero-Sutherland-Moser model is exactly solvable in
both the classical and the quantum regime\cite{Calog,Suth,Mos}.
This model is related to quantum spin chains with long range
interactions between the spins \cite{HalShas}, wave propagation in
stratified fluids \cite{CLP}, random matrix theory \cite{Suth,SLA}
and fractional statistics \cite{FStat}.
Remarkably, the quantum
solution is much easier to interpret, exhibiting a straightforward
analogy to the free fermion case. In a recent paper, Sutherland and
Campbell examined the classical system in the thermodynamic limit
and identified the excitations \cite{SC}. It was found that the classical
system has solitons, corresponding to a single particle running through
the rest of them, as well as small amplitude waves (phonons), identified
with holes. I shall derive here large amplitude wave
and soliton solutions of the classical system in the continuous limit,
where the particles form a ``fluid," and examine their correspondence to
the quantum states.

Consider a collection of particles of unit mass with the hamiltonian
\begin{equation}
H = \half \sum_{i=1}^N {\dot x}_i^2 + \sum_{i>j} {g \over ( x_i - x_j )^2}
\end{equation}
We shall be interested in the limit $N,L \to \infty$
with $N/L$ fixed. In this limit, the system can be described in terms of
a density field $\r (x)$ and a velocity field $\v (x)$.
At equilibrium, the particles will form
a regular lattice of spacing $a$ and density $\r_o = 1/a$. The particle
current is $J=\r \v$ and by particle conservation
\begin{equation}
\rd + \d J = \rd + \d (\r \v)=0
\end{equation}
where $\d = \d / \d x$. The kinetic energy of the system is
\begin{equation}
K=\int dx \half \r \v^2
\end{equation}
We can formally solve eq.~(2) for $\v$ to obtain $\v = -\d^{-1} \rd / \r$,
and the expression for the kinetic energy becomes
\begin{equation}
K = \int dx {(\d^{-1} \rd )^2 \over 2\r}
\end{equation}
This is exactly the kinetic term of the collective field hamiltonian
description of a many-body system \cite{BZ}. The potential energy can also be
expressed in terms of the density. The naive expression, however
is incorrect, the reason being that the interaction is singular at
coincidence points, and thus a substantial part of the potential
energy comes from nearest neighbors. The correct expression
requires a careful conversion of the discrete sum in terms of the
continuous fields. Alternatively, we can simply take the classical
limit ($\hbar \to 0$) of the quantum mechanical expression derived
in the collective field formulation \cite{AJL}. The result is
\begin{equation}
V = \int dx \left\{ {\pi^2 g \over 6} \r^3 - {g\over 2} \r \d \rt
+ {g\over 8} {(\d\r)^2 \over \r} \right\}
\end{equation}
where $\rt$ stands for the Hilbert transform:
\begin{equation}
\rt = \int dy ~P.P.{1\over x-y} ~\r(y)
\end{equation}

The dynamics of the system can be found by varying the lagrangian
$L=K-V +\mu \r$ with respect to $\r$. The chemical potential $\mu$
plays the role of a Lagrange multiplier ensuring that the total number of
particles remains constant. The resulting equations of motion are
\begin{equation}
\d^{-1}{\dot \v} + \half \v^2 +{\pi^2 g\over 2} \r^2 -g~\d\rt
+{g\over 8} \Bigl( {\d\r \over \r}\Bigr)^2 - g
{\d^2 \r \over 4\r} = \mu
\end{equation}
as well as the continuity equation.
By requiring that the static configuration $\v=0$, $\r = \r_o$ be a
solution, we obtain the value of the chemical potential
\begin{equation}
\mu = {\pi^2 g \over 2} \r_o^2
\end{equation}

For a localized constant profile configuration, propagating at speed
$\rv$, both $\r$ and $\v$ are functions of $x-\rv t$ only.
{}From the continuity equation we have
\begin{equation}
\d(\v \r - \rv \r ) = 0 ~~~~{\rm and~thus}~~~~ \v = {\r - \r_o \over
\r} \rv
\end{equation}
In the above, the integration constant is fixed by the boundary
condition that $\v \to 0$ at $x \to \pm \infty$, where $\r \to \r_o$.
Similarly, the eq. of motion becomes
\begin{equation}
{\rv^2 \over 2g} \Bigl({\r_o^2 \over \r^2} - 1 \Bigr) + {\pi^2 \over
2} (\r^2 - \r_o^2 ) = \d \rt -{1\over 8} \Bigl({\d\r \over \r}
\Bigr)^2 + {\d^2 \r \over 4\r}
\end{equation}
A solution of the above equation is
\begin{equation}
\rs = \r_o \left( 1 + {u \over (\pi \r_o x)^2 + u^2} \right)~,
{}~~~~ u = {\rv_s^2 \over \rv^2 - \rv_s^2}
\end{equation}
provided that $\rv > \rv_s$.

The above soliton carries particle number $Q$, momentum $P$
and energy $E$, defined as the extra amount over the static
solution $\r_o$. We find
\begin{equation}
Q = \int dx ~(\rs - \r_o) = 1
\end{equation}
\begin{equation}
P = \int dx ~\rs ~\v = \rv
\end{equation}
\begin{equation}
E = \int dx ~[K(\rs)+V(\rs) - V(\r_o)] = \half \rv^2
\end{equation}
We observe that the net particle number carried by the soliton
is 1, independently of its velocity; its momentum and energy are
also those of a free particle of unit mass moving at the soliton
velocity $\rv$. Therefore, the soliton can be exactly identified
with a particle excitation of the system. This is in agreement
with exact results drawn from the quantum theory, where particle
excitations always move faster than the sound \cite{SC}
The form of the above soliton, however,
is at odds with the results found in \cite{SC}. We suspect that the source of
the discrepancy is the truncation to a finite number of $x$-derivatives
of the form for the potential in \cite{SC}; this turns the equation to a
local one and gives the soliton an exponential decay, rather than the
inverse-square decay of the nonlocal equation.

A finite-amplitude periodic solution for the equations of motion is
\begin{equation}
\rw = \r_o + {1\over \l} \left( {1 \over \sqrt{\l^2 A^2 + 1} -
\l A \cos{2\pi x \over \l}} - 1 \right)
\end{equation}
where
\begin{equation}
\rv = \Bigl( \rv_s - {\pi \sqrt{g} \over \l} \Bigr)
\sqrt{1+{2 A^2 (\l \r_o - 1) \over \r_o^2 (1 + \sqrt{\l^2 A^2 +1} )}}
\end{equation}
$A$ is the amplitude of the wave and $l$ its wavelength. The
above equation is, therefore, the amplitude-dependent
dispersion relation in this nonlinear system. Note that in the
limit $\l \to \infty$ the above equations reduce to the single
soliton solution.

In summary, we have found exact soliton and wave solutions for the
CS system in the continuum limit. Certainly the above do
not exhaust the list of solutions; the general motion of the system
will be a nonlinear superposition of waves (or solitons).
It is instructive to put the above solutions into correspondence
with the quantum mechanical states. Consider $N$ particles in a space
of length $L$. The ground state of the system consists of a ``Luttinger
sea" in the pseudomomentum, with spacing between adjacent particles equal
to $2\pi \ell /L$ and ``Fermi level" $\pi \ell N/L$, where
$g=\ell (\ell -\hbar )$.
At the limit $\hbar \to 0$, $N,L \to \infty$, $N/L \to \r_o$, the
ground state becomes a continuous filled band with Fermi level $P_F =
\pi \sqrt{g} \r_o$.  A small amplitude wave, corresponding to a hole,
is a very small gap in the band. A soliton, corresponding to a particle
excitation, is a single particle peeled from the Fermi level and
placed some distance above. The generic finite amplitude wave corresponds
to a state with {\it two} continuous filled bands, of widths $P_1$ and
$P_2$ (with $P_1 + P_2 = 2 P_F$) and with a gap $G$ between them. These
are related to the wave parameters as
\begin{equation}
\l = {2\pi \sqrt{g} \over P_1 }
\end{equation}
\begin{equation}
\rv = {P_2 \over 2} \left({G \over \pi \sqrt{g} \r_o} + 1\right)
\end{equation}
Such a state can be visualized as arising either by successively
exciting single particles by the same constant momentum, until they
form a continuous band, or by gradually augmenting the gap of a hole,
until it becomes finite. This state can thus be thought of as either
a coherent state of solitons, or as a coherent
state of  phonons, their nonlinear
nature accounting for the change in profile as they accumulate.
Indeed, the soliton itself can be thought of as a superposition of
many phonons with very large wavenumber, and the phonon as a superposition
of many solitons just above the Fermi level. For the finite $N$ (finite
$L$) system the distinction between the two is fuzzy and in principle
only one kind of excitations need be considered as fundamental. Note,
further, that quantum mechanically the holes behave as particles with
fractional statistics of order $\hbar /\ell$ (meaning that $\ell/\hbar$
of them put together would form a fermion). At the classical limit
$\hbar \to 0$, thus, they become bosons, as they should be since
phonons obey no exclusion principle. Particles, on the other hand,
carry statistics of order $\ell/\hbar$. Thus in the classical limit
they become ``superfermions" meaning that no two of them can occupy
relatively nearby quantum states. This is consistent with the inverse
square repulsion between the classical particles. It should be noted
that a similar soliton solution exists in the fluid description of
this problem keeping the $\hbar$ corrections \cite{ABJ}, and it would
be intersting to find its interpretation.

We conclude by noting that the quantum mechanical problem separates
into two noninteracting chiral sectors, having to do with excitations
near either end of the Luttinger sea. (The two sectors mix nonperturbatively
when a number of particles of order $N$ is excited, depleting the sea.)
Therefore, the equation governing the continuum system should also
decompose into two nonmixing, first-order in time equations, one for
each sector. For the corresponding equation for free fermions this is
indeed the case \cite{Pol}. In fact, from the collective field description of
the system when only one chiral sector is present, we deduce that the chiral
equations are exactly of the Benjamin-Ono type \cite{AJL,MP}.
The exact field combinations in terms of which
this decomposition would be achieved, however, are not known and
constitute an open problem.

\section{EXCLUSION STATISTICS}

Statistics is an inherently quantum mechanical property of identical
particles which, as the name suggests, modifies the statistical
mechanical properties of large collections of such particles.
Motivated by properties of the Haldane-Shastry model (a lattice
version of the Calogero model), Haldane defined a generalized
{\it exclusion} statistics through the reduction of the Hilbert space
of additional particles in a system due to the ones already present
in the system \cite{Hal}. He proposed then the definition
\begin{equation}
g = -{\Delta d \over \Delta N}
\end{equation}
where $N$ is the number of particles in the system, $d$ is the
dimensionality of the single-particle Hilbert space, obtained by
holding the quantum numbers of $N-1$ particles fixed, and $\Delta
d$ and $\Delta N$ are their variation keeping the size and boundary
conditions of the system fixed. $g=0$ corresponds to bosons (no
exclusion) while $g=1$ corresponds to fermions, excluding a single
state for the remaining particles, the one they occupy.

On the basis of the above, Haldane proposed the combinatorial formula
for the number of many-body states of $N$ particles occupying
a group of $K$ states
\begin{equation}
M={[K-(g-1)(N-1)]! \over N! [K-g(N-1)-1]!}
\end{equation}
Starting from this, the thermodynamical properties of
particles with exclusion statistics can be derived \cite{Isa,Wu,Raj},
and this system has received a lot of recent attention
\cite{MuSh,VeOu,NaWi,DiPa}.

It is obvious from that exclusions statistics makes
sense only in a statistical sence, since $\Delta d$ and $M$ can
become fractional for $\Delta N =1$ or $N=1$. It is, nevertheless,
useful to attempt a {\it microscopic} realization and
interpretation of fractional exclusion statistics, and see what
it implies for the one-particle states.
Such a description has the obvious advantage
of being generalizable to {\it interacting}
particles, for which the notion of $d$ becomes hard to define.

The starting point will be the grand partition function for
exclusion-statistics particles in $K$ states
\begin{equation}
Z (K) = \sum_{N=0}^\infty M(K,N) x^N
\end{equation}
where we put $x=\exp(\mu-\varepsilon)/kT$ with $\mu$ the chemical
potential and assumed that all $K$ states are at the same energy
$\varepsilon$. In the statistical limit of large $K$, $Z(K)$
should be extensive. This introduces, then, the notion of a microscopic
description of the system in which the above $Z$ is the $K$-th power
of a single-state partition function. Each single level can be occupied
by any number of particles, but with an {\it a priori probability} $P_n$
for each occupancy $n$ independent of the temperature. We thus demand
\begin{equation}
Z(K,x) = \left( \sum_n P_n (K) x^n \right)^K =
\sum_{N=0}^\infty M(K,N) x^N
\end{equation}
for all $x$. The above probabilities must, in general, depend on $K$.
If, however, $P_n (K)$ assume some (finite) asymptotic values as $K$
goes to infinity (as they should for an extensive $Z$),
then the above microscopic partition function becomes
an accurate description in the statistical limit. This indeed happens,
and we obtain
\begin{equation}
P_n = \prod_{m=2}^n \Bigl( 1- {gn\over m}\Bigr)
\end{equation}

The most obvious feature of the above expressions is that,
unless $g=0,1$, they {\it always} become negative for some values
of $n$. Therefore, their interpretation as probabilities is
problematic. This is an inherent problem of fractional
$g$-statistics which cannot be rectified by, e.g., truncating
$M(K,N)$ to zero for $N>K/g$. The description of the
{\it statistical} system in terms of effective negative
microscopic probabilities is, nevertheless, accurate and useful.
Note, also, that the above $P_n$ never truncate to zero for
$n$ above some maximal value (unless $g=1$), unlike parafermions.

{}From the above, the single-level partition
function $Z(x) \equiv Z$ can be shown to satisfy
\begin{equation}
Z^g - Z^{g-1} = x
\end{equation}
The average occupation number $\bar n$ is expressed as
\begin{equation}
{\bar n} ={1\over Z} x \partial_x Z = x \partial_x W
\end{equation}
where $W= \ln Z$ is the thermodynamic potential (over $-kT$).
It can be shown that the above imply
\begin{equation}
(1-g{\bar n})^g [1-(g-1){\bar n}]^{1-g} = {\bar n} x^{-1}
\end{equation}
in accordance with the result of \cite{Isa,Wu,Raj}.

{}From the above we can easily calculate the low-temperature
expansion coefficients of the energy (or specific heat) in
terms of the temperature. Except for factors depending on
the dimensionality of space, these are \cite{NaWi}
\begin{equation}
C_n = \int_0^1  {dx \over x} (\ln x)^n [ (-1)^n {\bar n}(x)
+{\bar n}(x^{-1}) - {1\over g} ]
\end{equation}
where $1/g$ is the saturation density for $n$
at zero temperature and $\varepsilon < \mu$. The trick is to use
(25) and change variable of integration from $x$ to $W$. Since
by (24) $x$ is an explicit expression of $W$, we obtain the
explicit integral
\begin{eqnarray}
{C_n = (-1)^n \int_0^\infty \left\{ [gW+\ln(1-e^{-W})]^n \right.}
\cr
{\left. -(gW)^n \right\} dW}
\end{eqnarray}
The first few coefficients are
\begin{equation}
C_0 = 0,~~C_1 = {\pi^2 \over 6},~~ C_2 = -\zeta (3) \,(1+g) ,
{}~~{\rm etc.}~
\end{equation}
We see that $C_0 = 0$ for all $g$, as conjectured in \cite{NaWi}.
This expresses the fact that the ground state of the
many-body $g$-on system is nondegenerate, which is expected
to be a generic feature of particle systems. $C_1$ is also independent
of $g$. This result has first been conjectured by S. Isakov
and then numerically verified by J.~Myrheim. The first analytical derivation
was provided by D.~Arovas \cite{AIM}.  In fact, from (28)
we see that $C_n$ is a polynomial in $g$
of degree $n-1$, its $k$-th coefficient being
\begin{equation}
C_{n,k} = (-1)^n \left(\matrix{n \cr k} \right) \int_0^\infty
W^k [\ln(1-e^{-W})]^{n-k} dW
\end{equation}
Through a change of variables we can show that the above
coefficients obey
\begin{equation}
C_{n,k} = C_{n,n-k-1}
\end{equation}
so only half of the coefficients need be calculated.

The above relation points to a mapping between the small-$g$
and large-$g$ regions. Indeed, this is related to a duality
relation in these systems. It is easy to see that $Z$ satisfies:
\begin{equation}
Z^{-1} (g,x^{-g}) + Z^{-1} (g^{-1} ,x) = 1
\end{equation}
{}From the above relation, the duality relation
for the density is recovered \cite{Raj,NaWi}
\begin{equation}
g {\bar n}(g,x) + g^{-1} {\bar n}(g^{-1},
x^{-{1/ g}}) = 1
\end{equation}
We regard the formula for $Z$ (32) as more fundamental since it seems
to be more generic. For instance, parafermions of order $p=1/g$
are defined such that at most $p$ particles can be put per state
with probabilities 1. Thus
\begin{equation}
Z_{par} = 1+x+\cdots x^p = {1- x^{p+1} \over 1-x}
\end{equation}
from which we can write the generalized parafermionic partition
function $Z_{par} (g,x)$ by simply putting $p=1/g$ above.
It can be seen that $Z_{par}$ also satisfies (32) but not (33).

The free energy $W$ can be expressed as a power series in $x$
\begin{equation}
W = \sum_{n=1}^\infty {w_n \over n} x^n
\end{equation}
in terms of the ``connected" weights
\begin{equation}
w_1 = P1 ,~~ w_2 = 2 P_2 - P_1^2 ,~~ w_3 = 3P_3
- 3 P_1 P_2 + P_1^3
\end{equation}
etc. We find for $w_n$:
\begin{equation}
w_n = \prod_{m=1}^{n-1} \Bigl( 1 - {gn \over m} \Bigr)
\end{equation}
These are remarkably similar to $P_n$ (except for the range of $m$).
Notice that the $w_n$ are {\it not} probabilities, but rather
cluster coefficients. In fact, $w_n = 1$ for bosons and
$w_n = (-)^{n-1}$ for fermions. Also, $w_2 = 1-2g$ \cite{MuSh}.

{}From the above expressions for $w_n$ we can find a path integral
representation for the partition function of exclusion particles
in an arbitrary
external potential. We start from the usual euclidean path integral
with periodic time $\beta$ for $N$ particles with action the sum of
$N$ one-particle actions,
and sum over all particle numbers $N$ with appropriate chemical
potential weights. Since the particles are identical, we must
also sum over paths where particles have exchanged final positions,
with weights equal to the inverse symmetry factors of the
permutation to avoid overcounting (compare with Feynman diagrams).
Thus the path integral for each $N$ decomposes into sectors
labeled by the elements of the permutation group $Perm(N)$.
By the usual argument, the free energy will be given by the sum
of all connected path integrals. It is obvious that these are
the ones where the final positions of the particles are a
cyclic permutation of the original ones (since these are the
only elements of $Perm(N)$ that cannot be written as a product
of commuting elements). These have a symmetry factor of $1/N$
corresponding to cyclic relabelings of particle coordinates
(compare with the factors of $1/n$ included in (35)).
They really correspond to one particle wrapping $N$ times
around euclidean time. Thus, if we weight these configurations
with the extra factors $w_N$, as we have the right to do since
they belong to topologically distinct sectors, we will
reproduce the free energy of a distribution of $g$-statistics
particles on the energy levels of the one-body problem, that is
\begin{equation}
{\cal W}(\beta,\mu) = \sum_{N=1}^\infty e^{\mu N}
\int {w_N \over N} \prod_{n=1}^N
Dx_n (t_n ) e^{-S_E [ x_n (t_n ) ]}
\end{equation}
where $S_E$ is the one-particle euclidean action and the paths
obey the boundary conditions $x_n (\beta) = x_{n+1} (0)$,
$x_N (\beta) = x_0 (0)$. ($x$ can be in arbitrary dimensions.)
The partition function will be the
path integral over all (connected and disconnected) paths, with
appropriate symmetry factors and a factor of $w_n$ for each
connected $n$-particle component.

It is clear that the above path integral is not unitary, since
the weights $w_n$ are not phases, nor does it respect cluster
decomposition, since the $w_n$ do not provide true representations
of the permutation group (unlike the $g=0,1$ cases). This is
again a manifestation of the non-microscopic nature of exclusion
statistics. It does make sense, nevertheless, at the statistical
limit.

The above path-integral realization can be extended to other
statistics. E.g., for parafermions of order $p$ (with $p$
integer) the corresponding weigts $w_n$ are
\begin{equation}
w_n = -p ~~ {\rm for}~~n=0~{\rm mod} (p+1)~,~~1~~{\rm otherwise.}
\end{equation}
This representation is more economical than the one calling
for $p$ distinct flavors of fermions and projecting
into invariant states (singlets) under {\it Perm(p)}. The origin of
the apparent non-unitarity and breakdown of cluster decomposition
in the above integral for parafermions is clear: it is due to the
above projection, which must be inserted in the (unitary)
many-flavor path integral.

The possibility to define statistics through the choice of
the coefficients $w_n$ suggests other possible generalizations.
Perhaps the simplest one is to choose
\begin{equation}
w_n = (-\alpha)^{n-1}
\end{equation}
that is, one factor of $-\alpha$ for each unavoidable particle
crossing. This leads to the statistical distribution for the
average occupation number ${\bar n}$
\begin{equation}
{\bar n} = {1 \over e^{(\varepsilon - \mu)\beta} + \alpha}
\end{equation}
which is the simplest imaginable generalization of the Fermi
and Bose distribution, and was extensively analyzed in \cite{AcNa}.
The combinatorial formula for putting $N$ particles in $K$ states
for the above $\alpha$-statistics is
\begin{equation}
M = {K (K-\alpha ) (K-2\alpha ) \cdots (K-(N-1)\alpha )\over N!}
\end{equation}
This can be thought as a different realization of the exclusion
statistics idea: the first particle put in the system has $K$
states to choose, the next has $K-\alpha$ due to the presence of
the previous one an so on, and dividing by $N!$ avoids overcounting.
Fermions and bosons correspond to $\alpha=1$ and $\alpha=-1$
respectively, while $\alpha=0$ corresponds to Boltzmann statistics
(as is also clear from the path integral, in which no
configurations where particles have exchanged positions are
allowed, but factors of $1/N!$ are still included).
The corresponding single-level probabilities are
\begin{equation}
P_n = \prod_{m=1}^{n-1} {1-m\alpha \over 1+m}
\end{equation}
For $\alpha=1/p$ with $p$ integer (a fraction of a fermion),
the above probabilities are all positive for $n$ up to $p$
and vanish beyond that. For $\alpha<0$ all probabilities are
positive and nonzero. Thus, the above system has a bosonic
($\alpha <0$) and a fermionic ($\alpha>0$) sector, with
Boltzmann statistics as the separator. It is a plausible
alternative definition of exclusion statistics, due to (42),
and has many appealing features, not shared by the standard
(Haldane) exclusion statistics, such as positive probabilities,
a maximum single-level occupancy in accordance with the fraction
of a fermion that $\alpha$ represents, and analytic expressions
for all thermodynamic quantities. It would be interesting to find
a physical system in which these statistics are realized.

We conclude by pointing out that, once we have the path integral
we can easily extend the notion of exclusion statistics
to interacting particles: we simply replace the action
$\sum_n S_E [x_n ]$ by the full interacting $N$-particle action,
thus circumventing all difficulties with combinatorial formulae.
In the interacting case one has to work with the
full partition function, rather than the free energy,
since topologically disconnected diagrams are still dynamically
connected through the interactions and do not factorize.
Applications of the above on physical systems would be welcome.

\end{document}